\documentclass{article}
\usepackage{graphicx} 
\usepackage{url}
\usepackage{amsmath}
\usepackage{array}
\usepackage{booktabs}
\usepackage{longtable}
\usepackage{authblk}            
\usepackage{orcidlink}      
\usepackage{listings}
\usepackage{xcolor}
\usepackage{listings}

\title{Centrality in Collaboration: A Novel Algorithm for Social Partitioning Gradients in Community Detection for Multiple Oncology Clinical Trial Enrollments}
\author[1]{Benjamin Smith\orcidlink{0009-0007-2206-0177}}
\author[2]{Tyler Pittman\orcidlink{0000-0002-5013-6980}}
\author[1,2]{Wei Xu\orcidlink{0000-0002-0257-8856}}

\affil[1]{University of Toronto, Toronto, Canada}
\affil[2]{University Health Network, Toronto, Canada}

\date{} 
\begin{document}

\maketitle

\begin{abstract}
Patients at a comprehensive cancer center who do not achieve cure or remission following standard treatments often become candidates for clinical trials. Patients who participate in a clinical trial may be suitable for other studies. A key factor influencing patient enrollment in subsequent clinical trials is the structured collaboration between oncologists and most responsible physicians. Possible identification of these collaboration networks can be achieved through the analysis of patient movements between clinical trial intervention types with social network analysis and community detection algorithms. In the detection of oncologist working groups, the present study evaluates three community detection algorithms: Girvan-Newman, Louvain and an algorithm developed by the author. Girvan-Newman identifies each intervention as their own community, while Louvain groups interventions in a manner that is difficult to interpret. In contrast, the author's algorithm groups interventions in a way that is both intuitive and informative, with a gradient evident in social partitioning that is particularly useful for epidemiological research. This lays the groundwork for future subgroup analysis of clustered interventions.
\end{abstract}
\newpage
\section{Introduction}\label{introduction}

When cancer patients complete standard treatments, and have not responded with being cured or in remission, they become candidates for clinical trials. These clinical trials are regulated studies registered by Health Canada\footnote{For more information, see https://www.canada.ca/en/health-canada/services/clinical-trials.html}
as opposed to quality assurance studies\footnote{Quality assurance
  studies in the context of medical studies are studies which look at
  drugs which are already approved for use, but the goals are focused on
  other aspects of care such as drug delivery or quality of care.}.
Patients who qualify may have been screen failures for other trials, have
experienced progressive disease, or are receiving maintenance therapy and
have been referred to a clinical trial by their oncologist or most
responsible physician. Ground truth shows that collaboration networks
between oncologists is a primary factor for further engagement in
subsequent clinical trials by patients after completion of the given
clinical trial that they are enrolled in. A possible approach to
understanding the structure of these collaboration networks is through
use of social network analysis (SNA) and community detection algorithms.

Social network analysis examines individual entities and their
relationships among them. The data is represented as a ``graph'' where
individual entities are referred to as ``nodes'' and their relationships
between them as ``edges'', which may be directional if specified (see
Figure 1). A primary area of study in SNA is the analysis of
interconnectivity of nodes, called "communities" and identification of
clusters through the use of algorithms called "community detection
algorithms". Rostami et al\textsuperscript{1} (2023) note that there is
no specific model which describes exactly what a "community" is. Generally,
community detection algorithms employ specific optimization strategies
to partition a large-scale complex network into a set of disjoint and
compact subgroups, often (but not always) without prior knowledge
regarding the number of subgroups and their sizes. Rostami et al further
note that it is commonly acknowledged that there is no unique community
detection algorithm that can accommodate all kinds of graphs, because of
the inherent variability in network structures and their respective
objective(s).

\begin{figure}
\centering
\includegraphics[width=4.14212in,height=1.65018in]{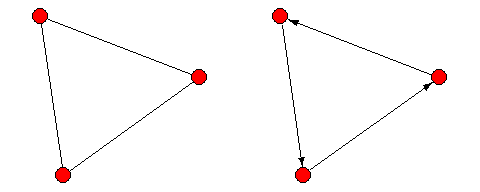}
\caption{Two simple graphs with directed and
undirected edges. Direction is noted by arrowheads at the end of the
edges.}
\end{figure}

Application of community detection algorithms with oncology clinical
trial data has been preformed in the past. Georgiev et
al\textsuperscript{2} (2011) applied the
Girvan-Newman\textsuperscript{3} (2002) algorithm and noted a lack of
cohesion among researchers who studied treatments for multiple myeloma.
Haq and Wang\textsuperscript{4} (2016) applied the Louvain algorithm (by
Blondel et al\textsuperscript{5} (2008)) to identify communities of
cancer patients with significantly different survival curves. The present study applies SNA, and compares multiple community detection algorithms to identify collaboration networks between oncologists through the interventions studied in clinical trials via enrollment data of patients in multiple, nonconcurrent clinical trials. Inspired by work
from Gorgiev et al (2011), Haq and Wang (2016), Ostovari and
Yu\textsuperscript{6} (2019) and Bissoyi and Patra\textsuperscript{7}
(2020) this research considers the Girvan-Newman and Louvain algorithms
and compares them to an author-developed algorithm, referred to as
"Smith-Pittman"\footnote{Named after the author and his co-supervisor,
  Tyler Pittman.}, to identify collaboration networks between clinical
trials classified by intervention.

\section{Materials and Methods}\label{materials-and-methods}

\subsection{The Data}\label{the-data}

The data is simulated oncology clinical trials. There were 2970 patients enrolled in 515 clinical trials involving 41 principal investigators. For the identification of collaboration networks
between oncologists, the analytic sample only consists of patients who
were enrolled in more than one clinical trial within the time period studied. 
The resulting analytic sample consists of 389 patients
enrolled in 288 clinical trials. Among these clinical trials, some
interventions can be classified into broader categories of
targeted therapies, or immunotherapy. This has been identified in the data
with ``T:'' and ``I:'' prefixes respectively. The clinical trials were
classified by intervention type, presenting as 16 distinct intervention types among
470 patient enrollments. With this classification, the patient referral
graph is constructed (see Figure 2).

The analysis is preformed with the \texttt{R} programming language, and
makes use of an extensive array of libraries and dependencies. The
primary libraries that were utilized include \texttt{igraph}, \texttt{tidyverse},
and \texttt{tidygraph}. For the complete script, please refer to the Appendix -
Program Syntax.
\begin{figure}
\centering
\includegraphics[width=5.88333in,height=4.65in]{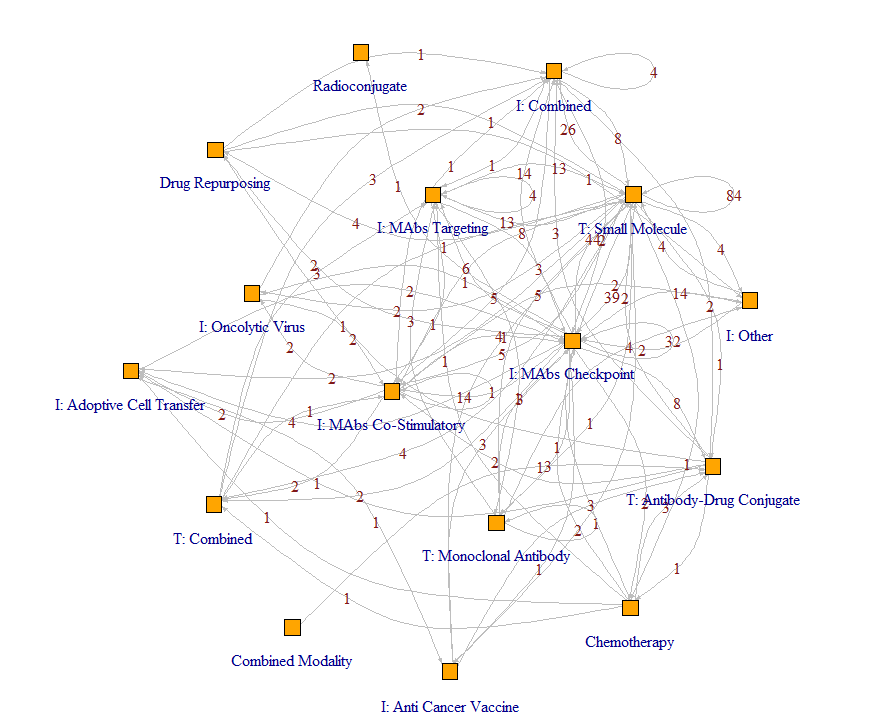}
\caption{Patient movement between clinical trials classified
by intervention type at PM. Nodes indicate the treatment type, and
labeled edges indicate the movement (subsequent enrollment) of patients between clinical trials in a given
intervention of the same type (self loop), or differing. Among the
clinical trials, some interventions can be classified into broader
categories consisting of targeted therapies or immunotherapy. This has
been identified in the data with ``T:'' and ``I:'' prefixes
respectively.}
\end{figure}

\subsection{Methods}\label{methods}

The goal of applying community detection algorithms with this data is to
identify oncologist working groups among treatment interventions, based on the movement
(incoming and outgoing referrals) of patients between the intervention types. These
movements in the network are understood through measures that are
considered by the community detection algorithms' optimization
strategies. While the Girvan-Newman, Louvain and Smith-Pittman
algorithms differ in their approaches to the identification of collaboration
networks, their identification strategies are based on the maximization of
modularity, \(Q\)- a measure that scores the degree of segregation
within a network through tightly connected communities or clusters (See
Newman\textsuperscript{8} (2006)).

The mathematical representation of modularity is defined in the
\texttt{igraph} R package\textsuperscript{9} (2006) as:

\[Q = \frac{1}{m}\sum_{i,j}^{}\left( A_{ij} - \frac{k_{i}^{\text{out}}k_{j}^{\text{in}}}{m} \right)\delta\left( c_{i},c_{j} \right)\]

Where \(m\) is the number of edges (patient movements), \(A_{ij}\) is
the number of connections shared by nodes \(i\) and \(j\) (movements
between interventions \(i\) and \(j\)), \(k_{i}^{\text{out}}\) and
\(k_{j}^{\text{in}}\) are the number of edges coming out from node \(i\)
and going into node \(j\) (patient movements from intervention \(i\) and
\(j\)) and \(\delta\left( c_{i},c_{j} \right)\) is an indicator variable
identifying if nodes \(i\) and \(j\) are connected - either directly or
through another node (if there is a patient movement between
interventions \(i\) and \(j\) either directly or through some other
intervention). For directed graphs, \(k_{i}^{\text{out}}\) and
\(k_{j}^{\text{in}}\) are simply the number of connected edges possessed
by nodes \(i\) and \(j\), respectfully. For a more comprehensive overview
modularity and other measures in social network analysis, see Newman
(2006), Wasserman and Faust\textsuperscript{10} (1994) and Latora et
al\textsuperscript{11} (2017).

\subsubsection{Girvan-Newman}\label{girvan-newman}

The Girvan-Newman algorithm is based on the evaluation of edges in a
social network through edge-betweenness centrality. Edge-betweenness
centrality is defined by Girvan and Newman (2002) as the number of
shortest paths that go through an edge in a graph, divided by the total
number of shortest paths between nodes \(i\) and \(j\). Each edge in a
graph has its own edge-betweenness centrality value.
The igraph (2006) documentation defines edge-betweenness centrality for an
edge \(e\) in a social network in mathematical terms as:

\[\sum_{i \neq j}^{}g_{iej}/g_{ij}\]

Where \(g_{ij}\) is the number of shortest paths between nodes \(i\) and
\(j\) (patient movements between interventions \(i\) and \(j\), either
directly or through some other intervention(s)), and \(g_{iej}\) is the
number of shortest paths which pass through edge \(e\). Figure 3
provides an illustration of a simple network, showing the edge with the
highest edge-betweenness centrality highlighted in red.
\begin{figure}
\centering
\includegraphics[width=3.03158in,height=2.02105in]{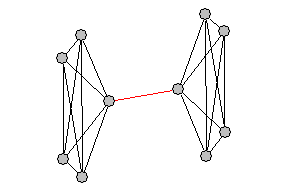}

\caption{A simple network demonstrating an edge with a high
edge-betweenness centrality, highlighted in red. The network consists of
two densely connected clusters, with the red edge serving as the sole
connection between them. This edge is crucial for communication between
the two clusters, as most of the shortest paths that connect nodes from
opposite clusters pass through it.}
\end{figure}

Edge betweenness can be calculated for directed and undirected edges. As
a result, the Girvan-Newman algorithm can be applied to directed or
undirected graphs without any transformations. The steps of the
Girvan-Newman algorithm are as follows:

\begin{enumerate}
\def\labelenumi{\arabic{enumi}.}
\item
  Calculate edge-betweenness centrality for all edges in the network.
\item
  Find the edge with the highest edge-betweenness centrality, and remove
  it from the network.
\item
  Recalculate edge-betweenness centrality for all remaining edges.
\item
  Repeat from step 2.
\end{enumerate}

Girvan-Newman can be utilized when the community structure is known, and will
classify nodes into a predetermined number of communities based on the
hierarchy produced by the algorithm (see Girvan and Newman 2002). When
the community structure is not known, modularity is evaluated after
each iteration of the algorithm. The grouping of nodes into distinct
communities is selected via modularity maximization.

\subsubsection{Louvain}\label{louvain}

The Louvain algorithm (by Blondel et al 2008) operates in two distinct
phases. (i) In the first phase, each node in the network is considered
as their own community, resulting in the initial partition with as many
communities as there are nodes. The algorithm then assesses the
potential modularity gain for each node \(i\) if it were to leave its
current community and join the community of node \(j\). After evaluating
the potential modularity gain across all communities, node \(i\) is
reassigned to the community of node \(j\), where the modularity increase
is maximized. The process is iteratively and sequentially applied for
all nodes, until no further improvement can be achieved. This first phase
stops when a local maximum of modularity is reached, meaning that no
subsequent node move can enhance modularity. (ii) The second phase
involves constructing a new network as represented by the communities identified
in the first phase. Links between nodes of the same community are viewed
as ``self-loops'' for the community in the new network. Once this
second phase is complete, the first phase of the algorithm can be
reapplied. The combination of these two phases is referred to as a
``pass''. The algorithm terminates when there is no other local maxima
in modularity to be achieved in subsequent passes.

A key limitation of the Louvain algorithm is that it is generally
programmed to work only with undirected graphs\footnote{Work on
  extending the Louvain algorithm to accommodate directed graphs has
  been an outstanding issue in the igraph developer community since 2015 (See:
  \url{https://github.com/igraph/igraph/issues/890}). However, Dugué and
  Perez\textsuperscript{12} (2022) have done some work on this.}. In
order to apply the Louvain algorithm to a directed graph, it must first
be converted to an undirected graph. Figure 4 is a reproduction of
Blodel et al's (2008) illustration of the algorithm.

\begin{figure}
\centering
\includegraphics[width=\linewidth]{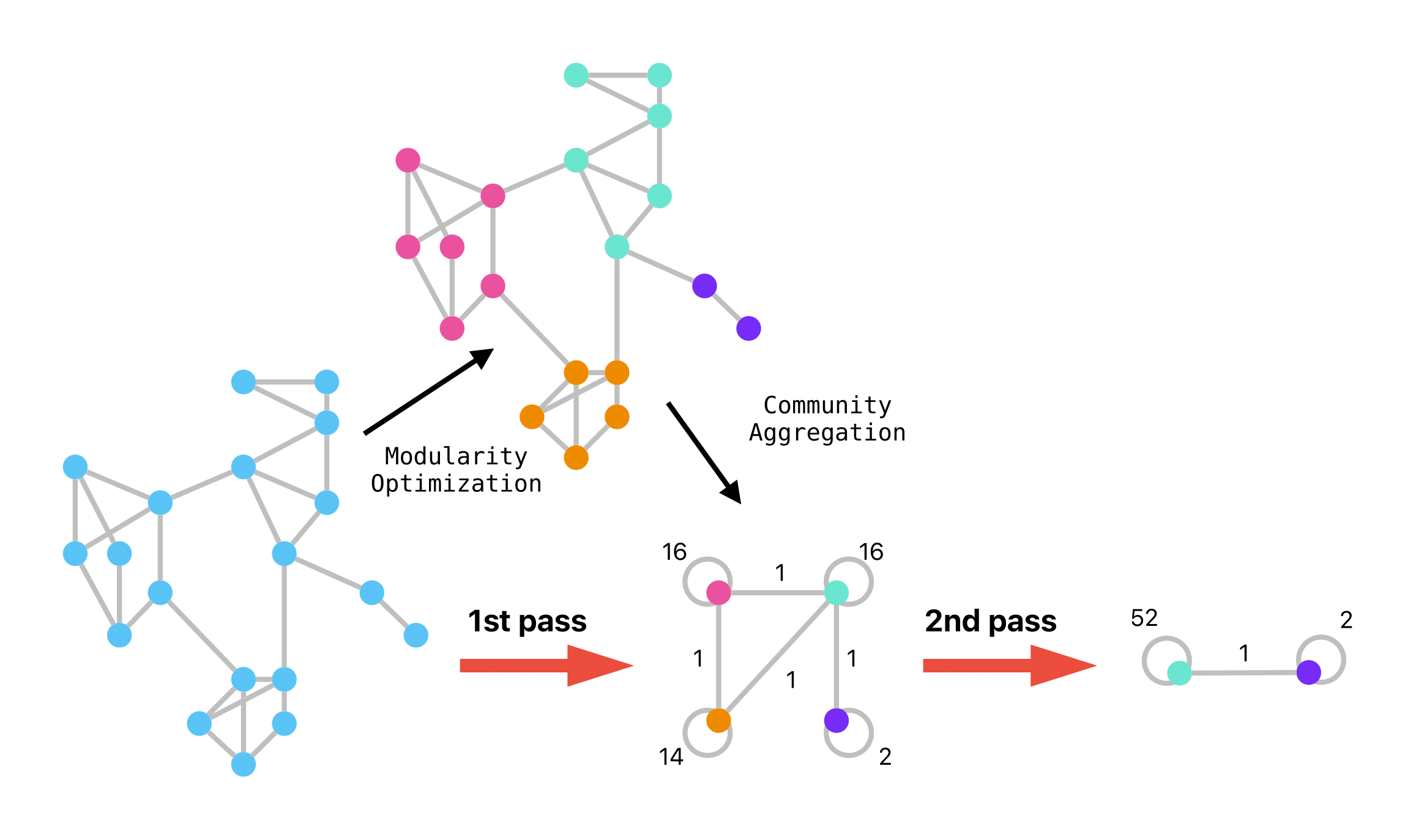}

\caption{Reproduced illustration of the Louvain algorithm
(originally designed by Blondel et al (2008)).}
\end{figure}

\subsubsection{Smith-Pittman}\label{smith-pittman}

The "Smith-Pittman" algorithm is a modification of the Girvan-Newman
algorithm, where degree centrality is considered. Degree centrality of a
node is simply defined as the number of connections a node has within a
given network (see figure 5). The algorithm proceeds through the
following steps:

\begin{enumerate}
\def\labelenumi{\arabic{enumi}.}
\item
  Calculate the degree centrality for each node, and the edge-betweenness
  centrality of all edges in the network.
\item
  Identify the subgraph associated with the node that has the highest
  degree centrality.
\item
  Remove the edge possessing the highest calculated edge-betweenness
  centrality.
\item
  Recalculate the degree centrality for all nodes, and the
  edge-betweenness centrality for the remaining edges in the network.
\item
  Repeat from step 2.
\end{enumerate}

Figure 6 provides a visual representation of this algorithm. Like
Girvan-Newman, the Smith-Pittman algorithm can be applied to both directed
and undirected graphs. Conceptually, the algorithm can be specified to
terminate once a predetermined number of communities have been
identified. However, its primary design is for use in an unsupervised
setting, where clusters are identified through the maximization of
modularity as evaluated after each iteration of the algorithm.

\begin{figure}
\centering
\includegraphics[width=2.02105in,height=2.02105in]{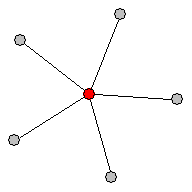}
\caption{A simple network highlighting node degree.
The center node (colored red) possesses the highest number of connections
and as a result possesses the highest degree and degree centrality
index.}
\end{figure}

\begin{figure}
\centering
\includegraphics[width=\linewidth]{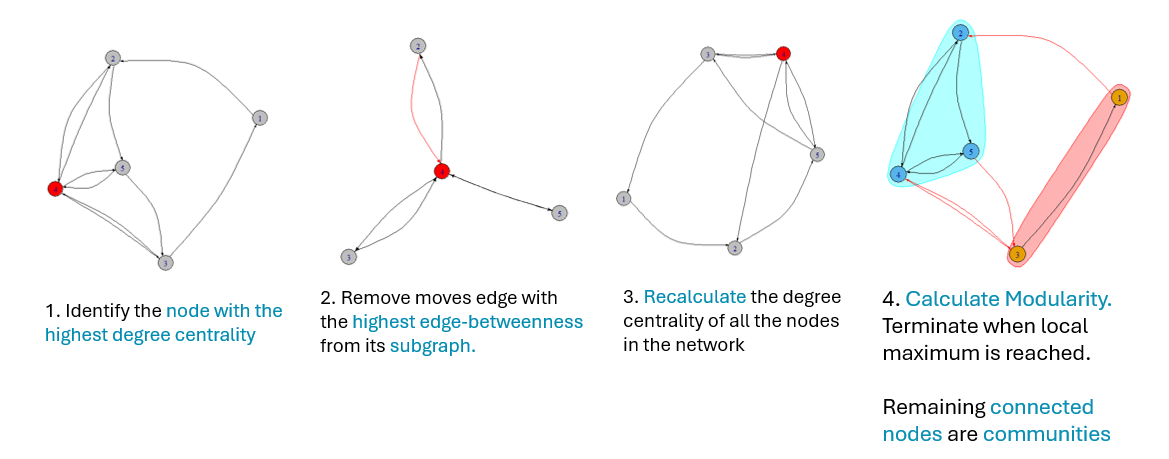}
\caption{Illustration of the Smith-Pittman algorithm.
Highlighted convex hulls denote the identification of distinct
communities.}
\end{figure}

\section{Results}\label{results}

Figures 7-9 show the communities identified by the algorithms, through
convex hulls highlighting the grouped interventions. Tables 1-3
show the grouping of interventions into communities, and the breakdown
by frequency of incoming and outgoing patient referrals for each treatment
intervention studied. Figure 7 demonstrates that the Girvan-Newman
algorithm identified each intervention as a separate community (\(Q =\)
0.044). This result is particularly uninformative, as it is equivalent
to not applying any community detection method to identify oncologist collaboration 
networks between the interventions. Figure 8 shows that the Louvain
algorithm groups interventions into four distinct working groups,
achieving the highest modularity score (\(Q =\) 0.177). However, the
underlying rationale and meaning behind these groupings remains unclear,
beyond the objective to cluster interventions as to maximize
modularity.

Figure 9 shows that the Smith-Pittman algorithm (\(Q =\) 0.08)
identified eight communities. Six of these communities consist of
individual interventions - namely T: Small Molecule, I:MAbs\footnote{Short
  for Monoclonal Antibodies.} Checkpoint, I:Combined, I:MAbs Targeting,
Combined Modality and Radioconjugate - while the remaining two
communities encompass multiple interventions. The interpretation of the
communities identified by the Smith-Pittman algorithm can be facilitated
by the degree of connectivity among the interventions within these
communities. Communities comprised of individual interventions either
have the highest or a substantial number of patient referrals, whether
incoming from or outgoing to other interventions, or they have the
least. Figure 10 illustrates the distribution of interventions by
patient referrals, ordered from smallest to largest, and highlights the
thresholds beyond which single intervention communities are positioned.
The interpretation of the communities identified by the Smith-Pittman
algorithm suggests the existence of both highly connected, and less
connected interventions, as well as broader groups corresponding to
typical intervention types - there is a gradient that is evident in social partitioning. This interpretation offers an intuitive
understanding related to the formation of collaboration networks being a
function of intervention ``popularity'' - i.e.~patient referrals
outgoing and incoming to and from other interventions.

\begin{figure}
\centering
\includegraphics[width=\linewidth]{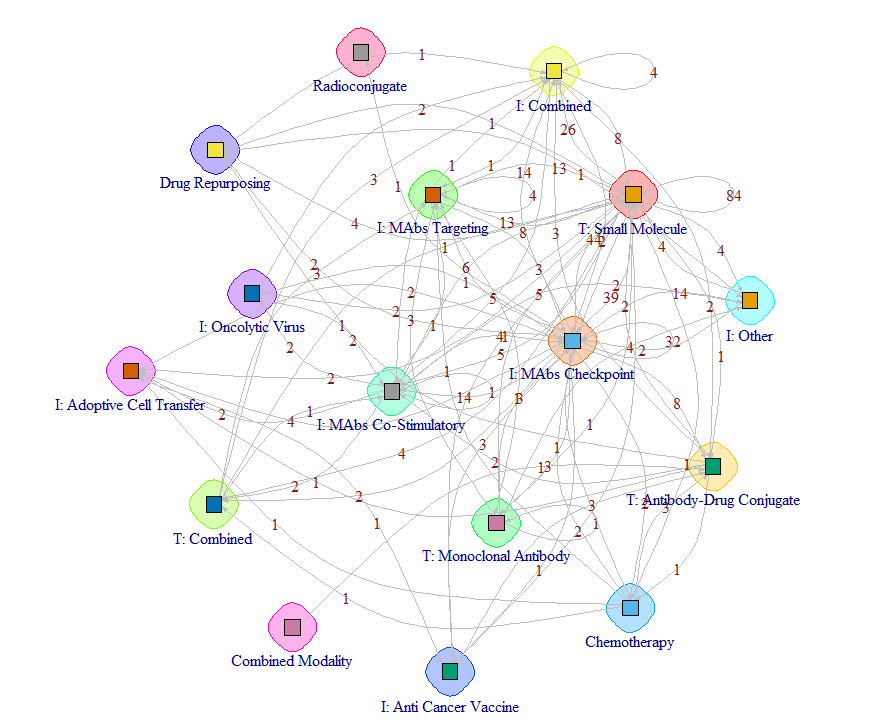}
\caption{Detected communities via Girvan-Newman with
modularity maximization. Sixteen distinct communities.}
\end{figure}

\begin{table}[htbp]
\centering
\begin{tabular}{lrrr}
\toprule
Intervention & Referrals In & Referrals Out & Total \\ 
\midrule
Chemotherapy & 4 & 10 & 14 \\ 
Combined Modality & 0 & 1 & 1 \\ 
Drug Repurposing & 7 & 3 & 10 \\ 
I: Adoptive Cell Transfer & 10 & 3 & 13 \\ 
I: Anti Cancer Vaccine & 4 & 7 & 11 \\ 
I: Combined & 54 & 22 & 76 \\ 
I: MAbs Checkpoint & 92 & 147 & 239 \\ 
I: MAbs Co-Stimulatory & 31 & 22 & 53 \\ 
I: MAbs Targeting & 31 & 22 & 53 \\ 
I: Oncolytic Virus & 4 & 5 & 9 \\ 
I: Other & 25 & 6 & 31 \\ 
Radioconjugate & 1 & 0 & 1 \\ 
T: Antibody-Drug Conjugate & 18 & 10 & 28 \\ 
T: Combined & 9 & 8 & 17 \\ 
T: Monoclonal Antibody & 6 & 16 & 22 \\ 
T: Small Molecule & 174 & 188 & 362 \\ 
\bottomrule
\end{tabular}
\caption{Girvan-Newman communities identified. Each intervention is their own community.}
\end{table}

\begin{figure}
\centering
\includegraphics[width=\linewidth]{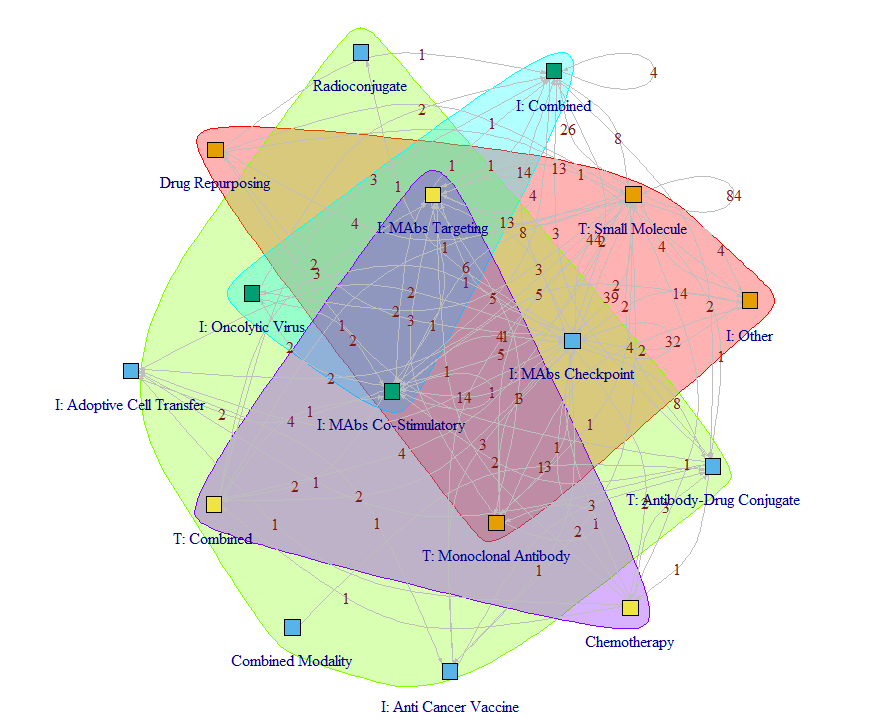}

\caption{Detected communities via Louvain algorithm with
modularity maximization. Four distinct communities.}
\end{figure}

\begin{table}[htbp]
\centering
\begin{tabular}{lrrr}
\toprule
Intervention & Referrals In & Referrals Out & Total \\ 
\midrule
\multicolumn{4}{l}{\textbf{Community: 1}} \\ 
\midrule
Drug Repurposing & 7 & 3 & 10 \\ 
I: Other & 25 & 6 & 31 \\ 
T: Monoclonal Antibody & 6 & 16 & 22 \\ 
T: Small Molecule & 174 & 188 & 362 \\ 
\midrule
\multicolumn{4}{l}{\textbf{Community: 2}} \\ 
\midrule
Combined Modality & 0 & 1 & 1 \\ 
I: Adoptive Cell Transfer & 10 & 3 & 13 \\ 
I: Anti Cancer Vaccine & 4 & 7 & 11 \\ 
I: MAbs Checkpoint & 92 & 147 & 239 \\ 
Radioconjugate & 1 & 0 & 1 \\ 
T: Antibody-Drug Conjugate & 18 & 10 & 28 \\ 
\midrule
\multicolumn{4}{l}{\textbf{Community: 3}} \\ 
\midrule
I: Combined & 54 & 22 & 76 \\ 
I: MAbs Co-Stimulatory & 31 & 22 & 53 \\ 
I: Oncolytic Virus & 4 & 5 & 9 \\ 
\midrule
\multicolumn{4}{l}{\textbf{Community: 4}} \\ 
\midrule
Chemotherapy & 4 & 10 & 14 \\ 
I: MAbs Targeting & 31 & 22 & 53 \\ 
T: Combined & 9 & 8 & 17 \\ 
\bottomrule
\end{tabular}
\caption{Louvain communities identified and grouped interventions}
\end{table}

\begin{figure}
\centering
\includegraphics[width=\linewidth]{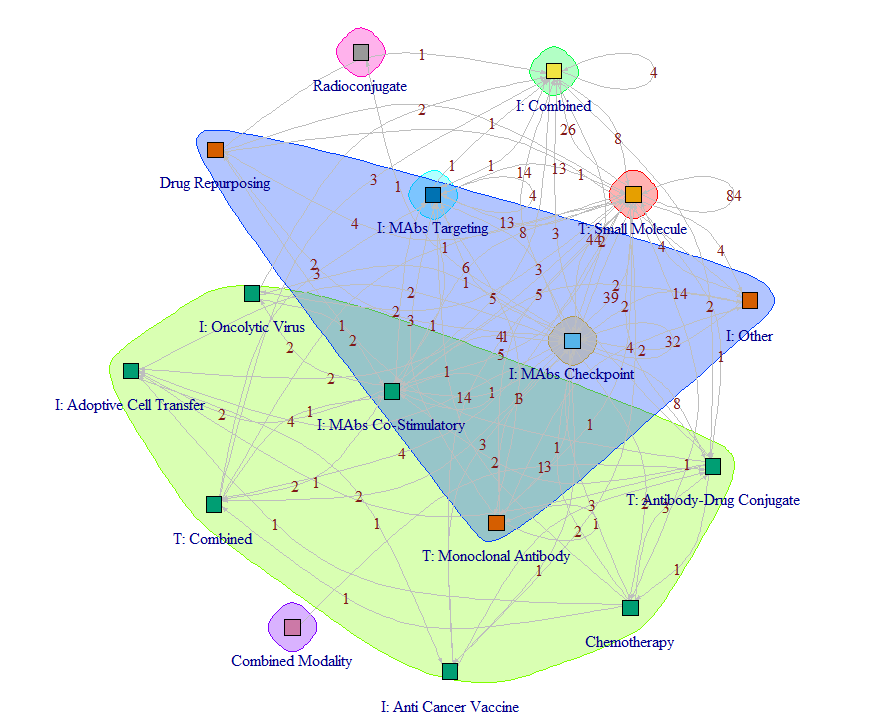}
\caption{Detected communities via Smith-Pittman
algorithm with modularity maximization. Eight distinct communities.}
\end{figure}

\begin{figure}
\centering
\includegraphics[width=\linewidth]{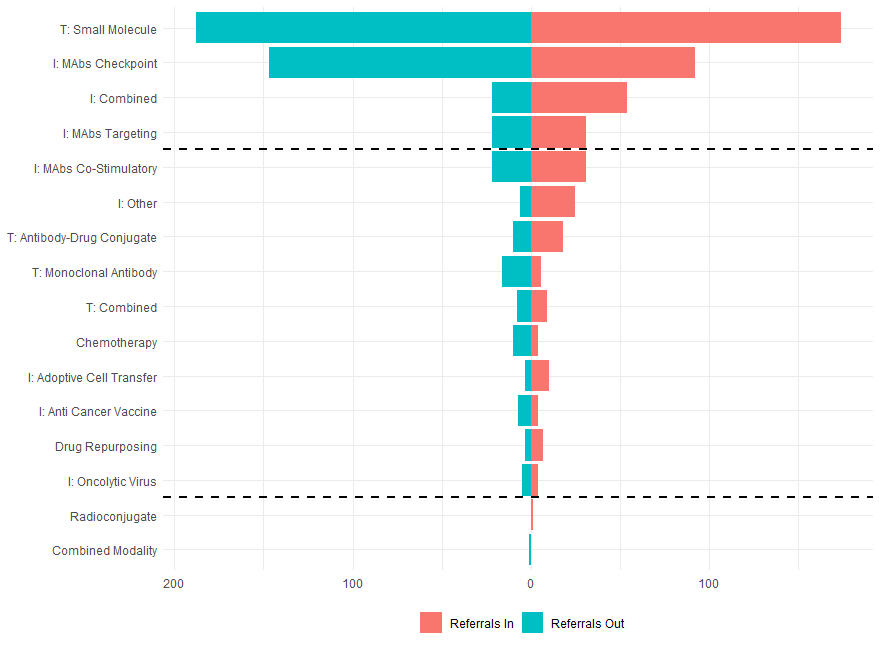}

\caption{Referral distribution among interventions.
Interventions outside the boundaries (T: Small Molecule, I:MAbs
Checkpoint, I: Combined, I:Mabs Targeting, Radioconjugate and Combined
Modality) are each identified as individual communities, while
interventions within them are identified as belonging to communities
consisting of multiple interventions.}
\end{figure}

\begin{table}[htbp]
\centering
\begin{tabular}{lrrr}
\toprule
Intervention & Referrals In & Referrals Out & Total \\ 
\midrule
\multicolumn{4}{l}{\textbf{Community: 1}} \\ 
\midrule
T: Small Molecule & 174 & 188 & 362 \\ 
\midrule
\multicolumn{4}{l}{\textbf{Community: 2}} \\ 
\midrule
I: MAbs Checkpoint & 92 & 147 & 239 \\ 
\midrule
\multicolumn{4}{l}{\textbf{Community: 3}} \\ 
\midrule
Chemotherapy & 4 & 10 & 14 \\ 
I: Adoptive Cell Transfer & 10 & 3 & 13 \\ 
I: Anti Cancer Vaccine & 4 & 7 & 11 \\ 
I: MAbs Co-Stimulatory & 31 & 22 & 53 \\ 
I: Oncolytic Virus & 4 & 5 & 9 \\ 
T: Antibody-Drug Conjugate & 18 & 10 & 28 \\ 
T: Combined & 9 & 8 & 17 \\ 
\midrule
\multicolumn{4}{l}{\textbf{Community: 4}} \\ 
\midrule
I: Combined & 54 & 22 & 76 \\ 
\midrule
\multicolumn{4}{l}{\textbf{Community: 5}} \\ 
\midrule
I: MAbs Targeting & 31 & 22 & 53 \\ 
\midrule
\multicolumn{4}{l}{\textbf{Community: 6}} \\ 
\midrule
Drug Repurposing & 7 & 3 & 10 \\ 
I: Other & 25 & 6 & 31 \\ 
T: Monoclonal Antibody & 6 & 16 & 22 \\ 
\midrule
\multicolumn{4}{l}{\textbf{Community: 7}} \\ 
\midrule
Combined Modality & 0 & 1 & 1 \\ 
\midrule
\multicolumn{4}{l}{\textbf{Community: 8}} \\ 
\midrule
Radioconjugate & 1 & 0 & 1 \\ 
\bottomrule
\end{tabular}
\caption{Smith-Pittman communities and identified and grouped interventions}
\end{table}

\section{Discussion}\label{discussion}

Where the Girvan-Newman algorithm failed to identify communities, the
Louvain and Smith-Pittman algorithms succeeded. A possible explanation
for this discrepancy lies in the nature of the data analyzed, which
includes patient referrals to clinical trials that investigate the same
intervention types as the clinical trials patients were previously
enrolled in. In graph theory, such referrals are represented as
``self loops'' and introduce complexity in the network. The
Girvan-Newman algorithm - whose original design was not for complex
networks - struggles in such contexts, leading to its failure to group
multiple interventions into communities based on modularity
maximization.

The Louvain algorithm successfully detected communities. However, the
resulting groups were difficult to interpret. This difficulty arises
because the Louvain algorithm bases its community selection purely on
modularity maximization, and does not consider the direction of patient
movements the underlying structural or functional significance of
particular interventions in the context of the network. The primary
advantage of the Louvain algorithm is its efficiency in preforming
community detection on large networks. It has been widely used in
applications such as the Twitter Social Network (Pujol et al.~2009)
which consisted of 2.4 million nodes and 38 million links, and mobile
phone network data (Greene et al.~2010) with 4 million nodes, 100
million links. These networks are orders of magnitude larger than the
patient referral network analyzed in this study, highlighting the
scalability of the Louvain algorithm. However, utility of such a
algorithm is limited in smaller, more specialized networks where the
interpretability and justification of communities identified is
important.

In contrast, the Smith-Pittman algorithm directly addresses connectivity
of interventions studied in the clinical trials, by incorporating degree
centrality and edge-betweenness centrality. This approach allows for the
identification of communities with a more ordered structure,
distinguishing between highly connected and minor interventions as they
reflect the relational dynamics in the network. The results from the
Smith-Pittman algorithm are promising, however, the results from this
analysis alone is insufficient to establish generalizability of the
algorithm. To fully assess its usefulness, a formal simulation study with
further application of the Smith-Pittman algorithm in diverse settings
is necessary. Additionally, the practical value of identified
communities will become evident when they are applied as grouping
variables in downstream analysis, such as outcome prediction or
intervention effectiveness studies.

Further research should focus on subgroup analysis, and exploring
extensions back to traditional statistical methods, such as regression
and survival analysis. This research can further validate the utility of
the identified communities, and the use of SNA and community detection
algorithms in clinical research settings. The results of the
Smith-Pittman algorithm lay the groundwork for these efforts, and
potentially offer a robust tool for community detection in social and
complex networks. Further work with the identified communities should
involve assessment of the impact of community structure on patient
outcomes, and identify if there are any structural inequities present in
clinical trial enrollments. This line of research can lead to the
identification of collaboration networks that improve patient care in
clinical settings.

\newpage
\section{References}\label{references}

1. Rostami, M., Oussalah, M., Berahmand, K. \& Farrahi, V. Community
Detection Algorithms in Healthcare Applications: A Systematic Review.
IEEE Access 11, 30247--30272 (2023).

2. Georgiev, H., Tsalatsanis, A., Kumar, A. \& Djulbegovic, B. Social
Network Analysis (SNA) of Research Programs In Multiple Myeloma (MM).
Blood 118, 3144--3144 (2011).

3. Girvan, M. \& Newman, M. E. J. Community structure in social and
biological networks. Proceedings of the National Academy of Sciences 99,
7821--7826 (2002).

4. Haq, N. \& Wang, Z. J. Community detection from genomic datasets
across human cancers. in 2016 IEEE Global Conference on Signal and
Information Processing (GlobalSIP) 1147--1150 (IEEE, 2016).
doi:10.1109/GlobalSIP.2016.7906021.

5. Blondel, V. D., Guillaume, J.-L., Lambiotte, R. \& Lefebvre, E. Fast
unfolding of communities in large networks. Journal of Statistical
Mechanics: Theory and Experiment 2008, P10008 (2008).

6. Ostovari, M. \& Yu, D. Impact of care provider network
characteristics on patient outcomes: Usage of social network analysis
and a multi-scale community detection. PLoS One 14, e0222016 (2019).

7. Bissoyi, S. \& Patra, M. R. Community Detection in a Patient-Centric
Social Network. in 171--182 (2021). doi:10.1007/978-981-15-7394-1\_17.

8. Newman, M. E. J. Modularity and community structure in networks.
Proceedings of the National Academy of Sciences 103, 8577--8582 (2006).

9. Csardi, G. \& Nepusz, T. The igraph software package for complex
network research. InterJournal, Complex Systems 1695 (2006).

10. Wasserman, S. \& Faust, K. Social Network Analysis. (Cambridge
University Press, 1994). doi:10.1017/CBO9780511815478.

11. Latora, V., Nicosia, V. \& Russo, G. Complex Networks. (Cambridge
University Press, 2017). doi:10.1017/9781316216002.

12. Dugué, N. \& Perez, A. Direction matters in complex networks: A
theoretical and applied study for greedy modularity optimization.
Physica A: Statistical Mechanics and its Applications 603, 127798
(2022).

\newpage

\section{Appendix - Program Syntax}\label{appendix---program-syntax}

\begin{lstlisting}[language=R]  % or Python, or whatever language
library(tidyverse)
library(tidygraph)
library(igraph)
library(ig.degree.betweenness) # Author developed methodology, pending public release
library(plyr) # for join_all
library(gt) # for tables
# Load R Data
real_df <- readRDS("path/to/data.rds")

real_df$New_Intervention_Name[real_df$New_Intervention_Name %in% c("Immunotherapy- MAbs-immunomodulatory-Checkpoint")] <- "I: MAbs Checkpoint";
real_df$New_Intervention_Name[real_df$New_Intervention_Name %in% c("Tageted therapy- antibody-drug conjugate")] <- "T: Antibody-Drug Conjugate";
real_df$New_Intervention_Name[real_df$New_Intervention_Name %in% c("Immunotherapy- MAbs-immunomodulatory-Co-Stimulatory")] <- "I: MAbs Co-Stimulatory";
real_df$New_Intervention_Name[real_df$New_Intervention_Name %in% c("Immunotherapy- Immuno + other investigational agent")] <- "I: Combined";
real_df$New_Intervention_Name[real_df$New_Intervention_Name %in% c("Targeted therapy - combined (small molecule + monoclonal antibody)")] <- "T: Combined";
real_df$New_Intervention_Name[real_df$New_Intervention_Name %in% c("Immunotherapy- MAbs- Tumour-targeting (includes immunoconjugates, naked MAbs)")] <- "I: MAbs Targeting";
real_df$New_Intervention_Name[real_df$New_Intervention_Name %in% c("Targeted therapy - small molecule")] <- "T: Small Molecule";
real_df$New_Intervention_Name[real_df$New_Intervention_Name %in% c("Immunotherapy- Other")] <- "I: Other";
real_df$New_Intervention_Name[real_df$New_Intervention_Name %in% c("Targeted therapy - monoclonal antibody")] <- "T: Monoclonal Antibody";
real_df$New_Intervention_Name[real_df$New_Intervention_Name %in% c("Immunotherapy- Adoptive Cell Transfer (e.g. TILS)")] <- "I: Adoptive Cell Transfer";
real_df$New_Intervention_Name[real_df$New_Intervention_Name %in% c("Immunotherapy- combined types")] <- "I: Combined";
real_df$New_Intervention_Name[real_df$New_Intervention_Name %in% c("Other - drug repurposing")] <- "Drug Repurposing";
real_df$New_Intervention_Name[real_df$New_Intervention_Name %in% c("Immunotherapy- Cytokines (eg. INFa, IL, Hematopoietic growth factors)")] <- "I: MAbs Co-Stimulatory";
real_df$New_Intervention_Name[real_df$New_Intervention_Name %in% c("Multiple- Biomarker Targeted")] <- "T: Combined";
real_df$New_Intervention_Name[real_df$New_Intervention_Name %in% c("Immunotherapy- Anti Cancer Vaccine- Peptide based vaccine")] <- "I: Anti Cancer Vaccine";
real_df$New_Intervention_Name[real_df$New_Intervention_Name %in% c("Chemotherapy")] <- "Chemotherapy";
real_df$New_Intervention_Name[real_df$New_Intervention_Name %in% c("Immunotherapy- Oncolytic Virus")] <- "I: Oncolytic Virus";
real_df$New_Intervention_Name[real_df$New_Intervention_Name %in% c("Combined modality (e.g chemoradiation, EBRT+Brachy)")] <- "Combined Modality";
real_df$New_Intervention_Name[real_df$New_Intervention_Name %in% c("Immunotherapy- Anti Cancer Vaccine- Gene Therapy (e.g DNA/RNA vaccines)")] <- "I: Anti Cancer Vaccine";
real_df$New_Intervention_Name[real_df$New_Intervention_Name %in% c("Other - radioconjugate")] <- "Radioconjugate";
real_df$New_Intervention_Name[real_df$New_Intervention_Name %in% c("Homonal Treatment")] <- "Drug Repurposing";


intervention_graph_real_directed<-  real_df |>
  dplyr::group_by(Subject_ID,Study_ID) |>
  dplyr::filter(dplyr::n() > 1) |>
  dplyr::distinct(pick(Subject_ID,Study_ID),.keep_all = TRUE) |>
  dplyr::ungroup() |>
  dplyr::filter(Subject_ID %in% names(table(Subject_ID))[table(Subject_ID) > 1]) |>
  dplyr::group_by(Subject_ID) |>
  dplyr::group_split() |>
  lapply(function(x) x |>
           dplyr::mutate(x, index = 1:nrow(x),
                         direction = ifelse(index%%2 == 1, "from","to"))) |>
  do.call(what = rbind) |>
  dplyr::select(Subject_ID, Study_ID, direction, New_Intervention_Name) |>
  tidyr::pivot_wider(
    id_cols = c(Subject_ID),
    names_from = direction,
    values_from = c(New_Intervention_Name, Study_ID)) |>
  dplyr::rename(from = New_Intervention_Name_from,
                to = New_Intervention_Name_to,
                Study_ID = Study_ID_from) |>
  tidyr::unnest(from) |>
  tidyr::unnest(to) |>
  tidyr::unnest(Study_ID) |>
  tidyr::unnest(Study_ID_to) |>
  dplyr::mutate(from = str_wrap(from, width = 30),
                to = str_wrap(to, width = 30)) |>
  #dplyr::group_by(from, to) |>
  #dplyr::count(name="Num_Patients") |>
  tidygraph::as_tbl_graph(directed = TRUE) |>
  igraph::as.igraph()


intervention_graph_real_undirected<-  real_df |>
  dplyr::group_by(Subject_ID) |>
  dplyr::filter(dplyr::n() > 1) |>
  dplyr::distinct(pick(Subject_ID,Study_ID),.keep_all = TRUE) |>
  dplyr::ungroup() |>
  dplyr::filter(Subject_ID %in% names(table(Subject_ID))[table(Subject_ID) > 1]) |>
  dplyr::group_by(Subject_ID) |>
  dplyr::group_split() |>
  lapply(function(x) x |>
           dplyr::mutate(x, index = 1:nrow(x),
                         direction = ifelse(index%%2 == 1, "from","to"))) |>
  do.call(what = rbind) |>
  dplyr::select(Subject_ID, Study_ID, direction, New_Intervention_Name) |>
  tidyr::pivot_wider(
    id_cols = c(Subject_ID),
    names_from = direction,
    values_from = c(New_Intervention_Name, Study_ID)) |>
  dplyr::rename(from = New_Intervention_Name_from,
                to = New_Intervention_Name_to,
                Study_ID = Study_ID_from) |>
  tidyr::unnest(from) |>
  tidyr::unnest(to) |>
  tidyr::unnest(Study_ID) |>
  tidyr::unnest(Study_ID_to) |>
  dplyr::mutate(from = str_wrap(from, width = 30),
                to = str_wrap(to, width = 30)) |>
  tidygraph::as_tbl_graph(directed = FALSE) |>
  igraph::as.igraph()


own_subj_mult_studies_check <- real_df |>
  dplyr::distinct(Subject_ID, Study_ID) |>
  dplyr::group_by(Subject_ID) |>
  dplyr::count(name="N_Studies") |>
  dplyr::filter(N_Studies > 1)
#389 participants enrolled in more than 1 study in 470 instances;



### Limit analysis to participants who enrolled in more than 1 clinical trial;


own <- real_df |>
  # Adding this line because Tyler has it as well. 
  dplyr::mutate(eligible = "eligible") |>
  dplyr::filter(Subject_ID %in% unique(own_subj_mult_studies_check$Subject_ID)) |>
  dplyr::select(
    "Subject_ID",
    "Study_ID",
    "Enrolled_Date_Time",
    "New_Intervention_Name",
    "PI_ID",
    "AE_Grade_3_Plus",
    "New_Intervention_Name",
    "eligible",
    "Age_40",
    "Age_65",
    "Baseline_AE",
    "New_Int_Name",
    "Phase",
    "Randomized",
    "Combination",
    "Sponsor_Type",
    "Disease_Site_Group"
  )



own_check <- own |>
  dplyr::select(Subject_ID, Study_ID, New_Intervention_Name, PI_ID) |>
  dplyr::filter(Subject_ID %in% unique(own_subj_mult_studies_check$Subject_ID)) |>
  dplyr::distinct(Subject_ID, Study_ID, New_Intervention_Name, PI_ID) |>
  dplyr::group_by(Subject_ID, Study_ID, New_Intervention_Name, PI_ID)

own_check <- as.data.frame(own_check)

###have to do New_Intervention_Name in here for correct department;
linkedDataStudies_0 <- own |>
  dplyr::distinct(New_Intervention_Name, Study_ID, Subject_ID, .keep_all = TRUE) |>
  dplyr::group_by(New_Intervention_Name, Study_ID) |>
  dplyr::count(name = "Num_Patients")


linkedDataStudies <- own |>
  dplyr::select(New_Intervention_Name) |>
  dplyr::group_by(New_Intervention_Name)

linkedDataPIs_0 <- own |>
  dplyr::distinct(PI_ID, New_Intervention_Name, Study_ID, Subject_ID) |>
  dplyr::group_by(PI_ID, New_Intervention_Name, Study_ID) |>
  dplyr::count(name = "Num_Patients")


linkedDataPIs <- own |>
  dplyr::distinct(PI_ID, New_Intervention_Name, Study_ID) |>
  dplyr::group_by(PI_ID)

linkedDataSubject_ID <- own |>
  dplyr::select(
    Subject_ID,
    Enrolled_Date_Time,
    Study_ID,
    New_Intervention_Name,
    New_Int_Name,
    eligible,
    Combination,
    Randomized,
    AE_Grade_3_Plus,
    Age_65
  ) |>
  dplyr::arrange(
    Subject_ID,
    Enrolled_Date_Time,
    Study_ID,
    New_Intervention_Name,
    New_Int_Name,
    eligible,
    desc(AE_Grade_3_Plus)
  ) |>
  dplyr::group_by(Subject_ID, Study_ID, New_Intervention_Name) |>
  dplyr::filter(row_number() == 1)

reach2=function(x){
  r=vector(length=vcount(x))
  for (i in 1:vcount(x)){
    n=neighborhood(x,2,nodes=i)
    ni=unlist(n)
    l=length(ni)
    r[i]=(l)/vcount(x)}
  r}

reach3=function(x){
  r=vector(length=vcount(x))
  for (i in 1:vcount(x)){
    n=neighborhood(x,3,nodes=i)
    ni=unlist(n)
    l=length(ni)
    r[i]=(l)/vcount(x)}
  r}

dwreach=function(x){
  distances=shortest.paths(x) #create matrix of geodesic distances
  diag(distances)=1 # replace the diagonal with 1s
  weights=1/distances # take the reciprocal of distances
  apply(weights,1,sum) # sum for each node (row)
}

#

fpntable <- table(own$Subject_ID);
otable <- table(own$PI_ID);
rtable <- table(own$New_Intervention_Name);
ownSmall <- own;
three_way_count  <- ownSmall |>
  dplyr::select(PI_ID, New_Intervention_Name, Study_ID, Subject_ID) |>
  dplyr::group_by(PI_ID, New_Intervention_Name, Study_ID, Subject_ID) |>
  dplyr::count(name="freq")
three_way_count <- as.data.frame(three_way_count);


edgelist <- cbind(three_way_count$PI_ID, three_way_count$New_Intervention_Name, three_way_count$Study_ID, three_way_count$Subject_ID, three_way_count$freq);

colnames(edgelist) <- c("PI_ID", "New_Intervention_Name", "Study_ID", "Subject_ID", "freq");
edgelist <- as.data.frame(edgelist);
linkedDataPIs_0 <- as.data.frame(linkedDataPIs_0);
linkedDataPIs <- as.data.frame(linkedDataPIs);
linkedDataStudies_0 <- as.data.frame(linkedDataStudies_0);
linkedDataStudies <- as.data.frame(linkedDataStudies);
linkedDataSubject_ID <- as.data.frame(linkedDataSubject_ID);

edgelist0 <- join_all(list(edgelist, linkedDataPIs_0, linkedDataStudies_0), by = c("New_Intervention_Name", "Study_ID"), type = "left", match = "first");
edgelist00 <- join_all(list(edgelist0, linkedDataSubject_ID), by=c("Subject_ID", "New_Intervention_Name", "Study_ID"), type="left", match = "first");
edgelist00_tibble <- as_tibble(edgelist00);


edgelist <- edgelist00[,colnames(edgelist00) %in% c("Subject_ID", "Study_ID", "Enrolled_Date_Time", "New_Intervention_Name", "PI_ID", "Num_Patients", "eligible", "Randomized", "Combination", "freq", "Department", "Enrolled_Date_Time", "Status_Change_Date_Time", "New_Int_Name", "AE_Grade_3_Plus", "Age_65")];

edgelist_count <- length(edgelist$Subject_ID);

n_pi <- length(unique(edgelist$PI_ID));
n_subjects <- length(unique(edgelist$Subject_ID));
n_studies <- length(unique(edgelist$New_Intervention_Name));
strat <- unique(eval(parse(text=paste("edgelist$", "eligible", sep=""))));
strat <- na.omit(strat);

edgelist <- edgelist[order(edgelist$Subject_ID, edgelist$Enrolled_Date_Time, edgelist$New_Intervention_Name),];
#which(is.na(eval(parse(text=paste("edgelist$", var[k], sep=""))))); #none, good check;

edgelist <- edgelist[,colnames(edgelist) %in% c("Subject_ID", "Study_ID", "New_Intervention_Name", "PI_ID", "Num_Patients", "freq", "eligible", "Randomized", "Combination", "Department", "Enrolled_Date_Time", "Status_Change_Date_Time", "New_Int_Name", "AE_Grade_3_Plus", "Age_65")];

edgelist$Subject_ID <- as.character(edgelist$Subject_ID);
edgelist$Study_ID <- as.character(edgelist$Study_ID);
edgelist$PI_ID <- as.character(edgelist$PI_ID);
edgelist$New_Intervention_Name <- as.character(edgelist$New_Intervention_Name);
edgelist$freq <- as.numeric(as.character(edgelist$freq));

edgelistPre_st <- edgelist

counterStrat <- length(strat)

st = strat;
strataCat = st;
edgelist <- edgelistPre_st |>
  filter(eval(parse(text="eligible")) == st)
edgelist <- as.data.frame(edgelist);
n_studies_strata <- length(unique(edgelist$New_Intervention_Name));

edgelist <- edgelist[order(edgelist$Subject_ID, edgelist$Enrolled_Date_Time, edgelist$New_Intervention_Name),];


###
### DO THIS FOR A SIMPLER DATASET TO CHECK CODE;
###
edgelist <- edgelist |>
  dplyr::arrange(Subject_ID, Enrolled_Date_Time) |>
  dplyr::group_by(Subject_ID) |>
  dplyr::mutate(order = row_number()) |>
  dplyr::mutate(from = Study_ID, 
         to = Study_ID,
         order_from = order, 
         order_to = order)

igraph0 <- edgelist |>
  dplyr::bind_rows(edgelist) |>
  dplyr::arrange(Subject_ID, Study_ID) |>
  dplyr::group_by(Subject_ID,Study_ID) |>
  dplyr::ungroup() |>
  dplyr::group_by(Subject_ID) |>
  dplyr::group_split() |>
  lapply(function(x) x |>
           dplyr::mutate(x, index = 1:nrow(x),
                         direction = ifelse(index%%2 == 1, "from","to"))) |>
  do.call(what = rbind) |>
  dplyr::select(Subject_ID, Study_ID, direction) |>
  tidyr::pivot_wider(
    id_cols = c(Subject_ID),
    names_from = direction,
    values_from = c(Study_ID)) |>
  tidyr::unnest(from, .drop=TRUE) |>
  tidyr::unnest(to, .drop=TRUE) 

igraph1 <- plyr::join_all(list(igraph0, edgelist[,c("Subject_ID", "from", "order_from")]), by=c("Subject_ID", "from"), type='left');

igraph2 <- plyr::join_all(list(igraph1, edgelist[,c("Subject_ID", "to", "order_to")]), by=c("Subject_ID", "to"), type='left');

igraph2 <- igraph2 |>
  dplyr::filter(order_from < order_to) |>
  dplyr::arrange(Subject_ID, order_from, order_to) |>
  dplyr::group_by(Subject_ID, from) |>
  dplyr::filter(row_number() == 1) |>
  dplyr::mutate(Study_ID_from = from,
                Study_ID_to = to) 

edgelist <- edgelist |>
  dplyr::arrange(Subject_ID, Enrolled_Date_Time) |>
  dplyr::group_by(Subject_ID) |>
  dplyr::mutate(order = row_number()) |>
  dplyr::mutate(New_Intervention_Name_from = New_Intervention_Name, 
                New_Intervention_Name_to = New_Intervention_Name)

igraph3 <- plyr::join_all(list(igraph2, edgelist[,c("Subject_ID", "from", "New_Intervention_Name_from")]), by=c("Subject_ID", "from"), type='left');
igraph4 <- plyr::join_all(list(igraph3, edgelist[,c("Subject_ID", "to", "New_Intervention_Name_to")]), by=c("Subject_ID", "to"), type='left');

igraph5 <- igraph4 |>
  dplyr::mutate(Study_ID_from = from,
                Study_ID = to,
                from = New_Intervention_Name_from,
                to = New_Intervention_Name_to) |>
  dplyr::select(-c("New_Intervention_Name_from", "New_Intervention_Name_to"))

igraph <- igraph5 |>
  dplyr::mutate(from = str_wrap(from, width = 30),
                to = str_wrap(to, width = 30)) |>
  tidygraph::as_tbl_graph(directed = TRUE) |>
  igraph::as.igraph()

e <- igraph::get.edgelist(igraph, names=FALSE);
l <- qgraph::qgraph.layout.fruchtermanreingold(e, vcount=vcount(igraph), area=30*(vcount(igraph)^2),repulse.rad=(vcount(igraph)^2.1));

# ########## Do this for a simpler graph just before plotting;
igraph_simplified <- igraph
E(igraph_simplified)$weight <- 1
igraph_simplified <- igraph::simplify(
  igraph_simplified,
  remove.multiple = T,
  remove.loops = F,
  edge.attr.comb = list(weight = "sum", "ignore")
)
E(igraph_simplified)$label <- E(igraph_simplified)$weight


# FOR VISUALS IN THIS REPORT

# Figure 1

g_directed <- graph(c(1, 2, 2, 3, 3, 1), directed = TRUE)
g_undirected <- as.undirected(g_directed)

V(g_directed)$color <- "red"
V(g_undirected)$color <- "red"
E(g_directed)$color <- "black"
E(g_undirected)$color <- "black"
set.seed(5208)
par(mfrow= c(1,2),mar=c(0,0,0,0)+.1)
plot(g_undirected,
     vertex.label = "",
     edge.arrow.size = 0.5,
     vertex.size = 20)
set.seed(5208)
plot(g_directed,
     vertex.label = "",
     edge.arrow.size = 0.5,
     vertex.size = 20)

# Figure 2



par(mar=c(0,0,0,0)+1)
plot(
  igraph_simplified,
  edge.label.color = "#801818",
  edge.label = E(igraph)$label,
  edge.label.cex = 1,
  edge.color = "grey",
  edge.arrow.size = 0.3,
  vertex.size = 5,
  vertex.shape = "square",
  vertex.color = "orange",
  vertex.label = V(igraph)$name,
  vertex.label.cex = 1.0,
  vertex.label.dist = 1.5,
  vertex.label.degree = pi / 2,
  edge.curved = TRUE,
  layout = l
)

# Figure 3


set.seed(5208)
par(mfrow= c(1,1),mar=c(0,0,0,0)+.1)
# Create two clusters
cluster1 <- sample(1:10, 5, replace = FALSE)
cluster2 <- sample(11:20, 5, replace = FALSE)
# Create edges within clusters
edges_within_cluster1 <- t(combn(cluster1, 2))
edges_within_cluster2 <- t(combn(cluster2, 2))
# Create edge connecting the clusters
edge_between_clusters <- matrix(c(sample(cluster1, 1), sample(cluster2, 1)), ncol = 2)
# Combine edges
edges <- rbind(edges_within_cluster1, edges_within_cluster2, edge_between_clusters)
# Create graph
g <- igraph::graph_from_edgelist(edges, directed = FALSE)
# Calculate betweenness centrality
betweenness_values <- igraph::edge_betweenness(g)
# Get the edge with the highest betweenness
max_betweenness_edge <- which.max(betweenness_values)
# Set edge color
igraph::E(g)$color <- "black"
igraph::E(g)[max_betweenness_edge]$color <- "red"
g <- igraph::induced_subgraph(g, which(igraph::degree(g) > 0))
# Plot the graph
plot(
  g,
  vertex.label = "",
  vertex.color = "grey",
  edge.curved = FALSE,
  edge.label = NA
)

# Figure 5

set.seed(5208)
par(mfrow= c(1,1),mar=c(0,0,0,0)+.1)

num_nodes <- 6

# Create an empty graph
g <- igraph::make_empty_graph(n = num_nodes)

# Add edges to connect all nodes to the central node (node 1)
for (i in 2:num_nodes) {
  g <- igraph::add_edges(g, c(1, i)) 
}

g |>
  igraph::as.undirected()|>
  plot(
    vertex.label="",
    vertex.color = ifelse(igraph::V(g)== 1, "red", "grey"),
    edge.color = "black"
  )


# Figures 7-10


# Putting this chunk here
gn_igraph <- igraph::cluster_edge_betweenness(igraph)

louvain_igraph <- igraph |>
  igraph::as.undirected() |> 
  igraph::cluster_louvain()

sp_igraph <- igraph |>
  ig.degree.betweenness::cluster_degree_betweenness()


# Figure 7

par(mar=c(0,0,0,0)+1)
plot(
  gn_igraph,
  igraph_simplified,
  edge.label.color = "#801818",
  edge.label = E(igraph)$label,
  edge.label.cex = 1,
  edge.color = "grey",
  edge.arrow.size = 0.3,
  vertex.size = 5,
  vertex.shape = "square",
  vertex.color = "orange",
  vertex.label = V(igraph)$name,
  vertex.label.cex = 1.0,
  vertex.label.dist = 1.5,
  vertex.label.degree = pi / 2,
  edge.curved = TRUE,
  layout = l
)

# Figure 8

par(mar=c(0,0,0,0)+1)
plot(
  louvain_igraph,
  igraph_simplified,
  edge.label.color = "#801818",
  edge.label = E(igraph)$label,
  edge.label.cex = 1,
  edge.color = "grey",
  edge.arrow.size = 0.3,
  vertex.size = 5,
  vertex.shape = "square",
  vertex.color = "orange",
  vertex.label = V(igraph)$name,
  vertex.label.cex = 1.0,
  vertex.label.dist = 1.5,
  vertex.label.degree = pi / 2,
  edge.curved = TRUE,
  layout = l
)

# Figure 9

par(mar=c(0,0,0,0)+1)
plot(
  sp_igraph,
  igraph_simplified,
  edge.label.color = "#801818",
  edge.label = E(igraph)$label,
  edge.label.cex = 1,
  edge.color = "grey",
  edge.arrow.size = 0.3,
  vertex.size = 5,
  vertex.shape = "square",
  vertex.color = "orange",
  vertex.label = V(igraph)$name,
  vertex.label.cex = 1.0,
  vertex.label.dist = 1.5,
  vertex.label.degree = pi / 2,
  edge.curved = TRUE,
  layout = l
)

# Figure 10

all_degree<- igraph::degree(igraph) |> 
  as.data.frame()|>
  tibble::rownames_to_column()|>
  dplyr::rename(degree=`igraph::degree(igraph)` ,
                study=rowname)

in_degree <- igraph::degree(igraph, mode = "in")|>
 as.data.frame()|>
 tibble::rownames_to_column()|>
 dplyr::rename(in_degree=`igraph::degree(igraph, mode = "in")` ,
               study=rowname)

out_degree <- igraph::degree(igraph, mode = "out") |>
   as.data.frame()|>
  tibble::rownames_to_column()|>
  dplyr::rename(out_degree=`igraph::degree(igraph, mode = "out")` ,
                study=rowname)

degree_df <- merge(in_degree,
                   out_degree)|>
  merge(all_degree)|>
  dplyr::mutate(in_degree = -in_degree)|>
  tidyr::pivot_longer(cols = c(in_degree,out_degree))

ggplot(degree_df,
       mapping = aes(y =reorder(study, degree), x = -value, fill = name))+
  theme_minimal()+
  geom_col()+
  geom_hline(yintercept = 2.5,linetype='dashed',lwd=1)+
  geom_hline(yintercept = 12.5,linetype='dashed',lwd=1)+
  theme(axis.title.y = element_blank(),
        legend.title = element_blank(),
        legend.position = "bottom",
        axis.title.x = element_blank())+
  scale_fill_manual(labels = c("Referrals In", "Referrals Out"), values = scales::hue_pal()(2))+
  scale_x_continuous(labels = abs)

# Tables

# Table 1

gn_df <- data.frame(
  Intervention = igraph::V(igraph)$name,
  "Patient Refferalls: In" = igraph::degree(igraph,mode="in"),
  "Patient Referrals: Out" = igraph::degree(igraph, mode="out"),
  "Total Patient Refferals" =  igraph::degree(igraph, mode="total"),
  row.names = NULL,
  check.names = FALSE
) |>
  dplyr::group_by(Intervention) |> 
  dplyr::summarise(
    `Refferalls In` = sum(`Patient Refferalls: In`),
    `Referrals Out` = sum(`Patient Referrals: Out`),
    `Total` = sum(`Total Patient Refferals`)
  )

gt::gt(gn_df)|>
  gt::tab_header("Table 1: Girvan-Newman communities identified. Each intervention is their own community.")|>
  gt::cols_width(
    Intervention ~ gt::pct(40),
    `Refferalls In` ~ gt::pct(15),
    `Referrals Out`  ~ gt::pct(20),
    `Total` ~ gt::pct(15)
    ) |>
  gt::tab_options(table.font.size=42)



# Table 2

louvain_df <- data.frame(
  Intervention = igraph::V(igraph)$name,
  Community = paste0("Community: ", igraph::membership(louvain_igraph)|> as.vector()),
  "Patient Refferalls: In" = igraph::degree(igraph,mode="in"),
  "Patient Referrals: Out" = igraph::degree(igraph, mode="out"),
  "Total Patient Refferals" =  igraph::degree(igraph, mode="total"),
  row.names = NULL,
  check.names = FALSE
)

louvain_df |>
  dplyr::group_by(Community,Intervention) |> 
  dplyr::summarise(
    `Refferalls In` = sum(`Patient Refferalls: In`),
    `Referrals Out` = sum(`Patient Referrals: Out`),
    `Total` = sum(`Total Patient Refferals`)
  )|>
  gt::gt()|>
  gt::tab_header("Table 2: Louvain communities identified and grouped interventions.")|>
  gt::cols_width(
    Intervention ~ gt::pct(40),
    `Refferalls In` ~ gt::pct(15),
    `Referrals Out`  ~ gt::pct(20),
    `Total` ~ gt::pct(15)
  )|>
  gt::tab_options(table.font.size=42)



# Table 3

sp_df <- data.frame(
  Intervention = igraph::V(igraph)$name,
  Community = paste0("Community: ", igraph::membership(sp_igraph)|> as.vector()),
  "Patient Refferalls: In" = igraph::degree(igraph,mode="in"),
  "Patient Referrals: Out" = igraph::degree(igraph, mode="out"),
  "Total Patient Refferals" =  igraph::degree(igraph, mode="total"),
  row.names = NULL,
  check.names = FALSE
)

sp_df |>
  dplyr::group_by(Community,Intervention) |> 
  dplyr::summarise(
    `Refferalls In` = sum(`Patient Refferalls: In`),
    `Referrals Out` = sum(`Patient Referrals: Out`),
    `Total` = sum(`Total Patient Refferals`)
  )|>
  gt::gt()|>
  gt::tab_header("Table 3: Smith-Pittman communities and identified and grouped interventions.")|>
  gt::cols_width(
    Intervention ~ gt::pct(40),
    `Refferalls In` ~ gt::pct(15),
    `Referrals Out`  ~ gt::pct(20),
    `Total` ~ gt::pct(15)
  )
\end{lstlisting}

\end{document}